\documentclass[conference, a4paper]{IEEEtran}
\IEEEoverridecommandlockouts

\ifCLASSINFOpdf
  \usepackage[pdftex]{graphicx}
  \DeclareGraphicsExtensions{.pdf,.jpeg,.png}
\else
\fi

\usepackage{cite}
\usepackage{amsmath,amssymb,amsfonts}

\usepackage{graphicx}
\usepackage{textcomp}
\usepackage{url}
\usepackage[left=1.4cm, right=1.4cm, top=1.7cm]{geometry}
\usepackage{caption}
\usepackage{anyfontsize}
\usepackage{multirow}
\usepackage{array}
\usepackage{algorithmicx}
\usepackage{algorithm} 
\usepackage{algpseudocode}
\usepackage{makecell}

\makeatletter
\newcommand*\titleheader[1]{\gdef\@titleheader{#1}}
\AtBeginDocument{%
\let\st@red@title\@title
\def\@title{%
\bgroup\normalfont\normalsize\centering\@titleheader\par\egroup
\vskip0.2em\st@red@title}
}
\makeatother

\makeatletter
\renewcommand{\fnum@figure}{Figure \thefigure}
\makeatother

\title{ {Robust Management of Airport Security Queues Considering Passenger Non-compliance with Chance-Constrained Optimization} \\
\vspace{0.5cm}
}

\titleheader{First US-Europe Air Transportation Research and Development Symposium (ATRDS2025)}

\author{\IEEEauthorblockN{Shangqing Cao, Aparimit Kasliwal, \\ Huangyi Zheng, Masoud Reihanifar, \\ Mark Hansen}
\IEEEauthorblockA{Department of Civil and Environmental Engineering \\
University of California, Berkeley \\
Berkeley, USA \\
caoalbert@berkeley.edu}
\and
\IEEEauthorblockN{Francesc Robusté}
\IEEEauthorblockA{BIT - Barcelona Innovative Transportation\\
Civil and Environmental Engineering Department\\
Polytechnic University of Catalonia (UPC) - BarcelonaTech\\
Barcelona, Spain}  
}

\IEEEaftertitletext{\vspace{-1\baselineskip}}

\begin{document}

\maketitle

\noindent \begin{abstract}
The long waiting time at airport security has become an emergent issue as demand for air travel continues to grow. Not only does queuing at security cause passengers to miss their flights, but also reduce the amount of time passengers spend at the airport post-security, potentially leading to less revenue for the airport operator. One of the key issues to address to reduce waiting time is the management of arrival priority. As passengers on later flights can arrive before passengers on earlier flights, the security system does not always process passengers in the order of the degree of urgency. In this paper, we propose a chance-constrained optimization model that decides in which time slot passengers should be recommended to arrive. We use chance constraints to obtain solutions that take the uncertainty in passenger non-compliance into account. The experimental results, based on a sample day of flight schedules at the Barcelona airport, show a reduction of 85\% in the total waiting time. Compared to the deterministic case, in which passengers are assumed to fully comply with the recommendations, we see a 30\% increase in the reduction of the total waiting time. This highlights the importance of considering variation in passenger compliance in the management of airport security queues. 

\end{abstract}

\vspace{0.3cm}


\section{Introduction}

Efficient management of passenger flow at airports is critical to minimize delays and ensure smooth operations. Passengers' arrival at the airport for catching their departing flights is the initial step in the process, any potential delays to which directly lead to congestion at security checkpoints, which further propagate to delays in aircraft departures and may even potentially result in missed flights. 

Airports around the world are facing a surge in traffic. The volume of passengers not only bounced back from the downturn caused by COVID-19, but reached a record number in 2024 globally \cite{iata}. The year-over-year increase in domestic and international traffic reached 5.7\% and 13.6\% \cite{iata}. Rapid increase in demand poses significant challenges to the infrastructure. The processing capacity of security is at the center of the stage. Airports such as Miami International Airport (MIA), Los Angeles International Airport (LAX), and John F. Kennedy International Airport (JFK) reported an average waiting time in peak periods of more than 54 minutes during the American winter holiday season in 2024 \cite{fox}. Airports such as St. Louis Lambert International Airport (STL), Austin-Bergstrom International Airport (AUS), and Orlando International Airport (MCO) reported an average waiting time of over 15 minutes throughout the year in 2024 \cite{bounce}. European airports, such as Hamburg Airport (HAM) and Stuttgart Airport (STR), also report an average highest peak wait time of 30 minutes and 20 minutes, respectively \cite{eu}. Even before COVID-19, a survey reported that long airport security queues caused one in seven passengers to miss a flight in the last 12 months \cite{c1}. The waiting time at security also plays an important role in changing passengers' perception of the airport. A study at Gimpol International Airport (GMP) found that perceived security waiting time has a statistically positive relationship with perceived boredom and a statistically negative relationship with acceptability \cite{gmp}. 

One of the key challenges in managing security checkpoints is the variability in passenger arrival times. A potential solution involves assigning specific time slots for airport arrivals, strategically aligned with flight departure schedules to optimize queue efficiency. However, this approach is inherently sensitive to variations in compliance rates, as not all passengers strictly adhere to the assigned arrival times. Differences in individual behavior and scheduling preferences introduce uncertainty, reducing the robustness of such an optimized system. 

Passenger arrival behavior varies according to individual preferences and flight characteristics, leading to differences in adherence to assigned arrival slots. Taking into account these variations in compliance rates, we develop a more robust framework using chance-constrained optimization to manage passenger flow and optimize airport security queues for efficiency. Our preliminary results indicate that accounting for compliance variability leads to reduced queue waiting times compared to both the conventional first-come-first-serve (FCFS) queuing system and a deterministic scheduling model that assumes complete compliance. This work contributes to the broader goal of improving airport security queue efficiency by integrating passenger behavior dynamics into strategic scheduling frameworks.

\section{Literature Review}

Various mathematical frameworks have been employed to optimize passenger flow within airport systems. Queuing theory has been widely used to model and analyze the dynamics of security checkpoints, with models that characterize the state of the security system through passenger service rate, the number of available security lanes, and the passenger arrival rate \cite{gilliam}. The existing management strategies can be categorized into two: changing service rates or changing arrival rates. For example, the introduction of priority lanes or optimizing security staffing are strategies that actively manage service rates so that more passengers can be accommodated during peak operating hours \cite{brun} \cite{b9} \cite{scozzaro}. On the other hand, the concept of virtual queuing, through which a passenger can enter the security queue without physically being present, induces changes in arrival rates to better match the system's service rate \cite{lange}. On a more microscopic level, designs such as lane assignment, changes the passenger arrival rate to each individual lane to match the arrival rate with the service rate \cite{marshall}. Both classes of methods have shown effectiveness in reducing security waiting time and increasing system output.


The problem of managing passenger flow at airport security checkpoints shares interesting similarities with the classical \textit{assignment problem}, which is widely studied in the optimization literature. The assignment problem involves the allocation of a set of resources to tasks in an optimal manner, often with constraints on availability and efficiency. In the context of airport security, assigning passengers to specific arrival time slots can be framed as an optimization problem, where time slots represent limited resources and the objective is to minimize waiting times while ensuring timely passage through security. However, a major limitation of deterministic models is their reliance on perfect compliance assumptions, which may not hold in real-world scenarios. Studies in operations research have explored variations of assignment problems, including robust and stochastic formulations, to handle uncertainty in resource availability and compliance rates \cite{b7} \cite{b6}. The use of chance-constraint, a formulation that incorporates probabilistic instead of deterministic constraints, has been widely adopted in optimization problems in transportation \cite{chen} \cite{wang}. These approaches provide a foundation for developing scheduling strategies that account for variability in passenger behavior.

It is crucial to consider passenger compliance and variability in passenger behaviors when designing effective management strategies for security systems. The reduction in the average waiting time has been found to be positive in relation to the participation rate in the control scheme \cite{lange}. Existing management strategies that focus on changing the arrival rates of passengers also assume that passengers do not renege. Therefore, there exists a gap in the current literature on designing control schemes that consider varying compliance patterns and participation rates.

In this work, we propose and test a control scheme that focuses on changing the arrival rates of passengers to reduce the waiting time at security. We extend the classical assignment framework by incorporating compliance variability into the optimization model. Our approach considers passengers’ adherence to their assigned time slots and accounts for deviations that may arise due to model human behavior or external factors. By introducing a chance-constrained model, we aim to enhance the resilience of scheduling strategies against unpredictable passenger behavior. Preliminary results indicate that incorporating compliance variability leads to significant reductions in average waiting times compared to both traditional FCFS queuing and deterministic scheduling models. Furthermore, our results suggest that optimizing assignment strategies under realistic behavioral constraints improves overall system efficiency without imposing unrealistic constraints on passengers.

The remainder of the paper is organized as follows. Section \ref{sec:math} details the key mathematical formulation of this work, including the chance-constrained optimization model and two baseline models for comparison: the queuing model and the deterministic optimization model. Section \ref{sec:exp} describes the numerical experiments conducted to evaluate the effectiveness of the proposed model in managing the congestion in security queues and reducing the waiting time. Section \ref{sec:sen_ana} presents a sensitivity analysis on a set of parameters in the chance-constrained optimization. 

\section{Mathematical Models}
\label{sec:math}

We formulate the security management problem as a second-order cone program with chance constraints, in which each passenger is recommended an arrival time slot at security checkpoints. By distributing the number of passenger arrivals over many different time slots and prioritizing passengers whose arrival at security is closer to the departure time of their flights, we can reduce the total waiting time at security while ensuring that no passenger misses the flight. The issue, however, is that passengers may not fully comply with the recommended time of arrival. Therefore, in designing the control policy, one must consider the varying level of compliance rate. In this section, we first establish a queuing system to quantify the delay incurred at security checkpoints and then present two optimization models, one deterministic and one with chance constraints, for the management of security queues.

\subsection{Queuing Model and Metric}
\label{sec:queuing}

An excess of waiting time is created when the number of passengers arriving at security checkpoints exceeds the processing capacity of the security system. We set up the following FCFS queuing model to quantify the total amount of delay incurred without any control in place. We introduce the following notation:
\begin{itemize}
\item $a(t)$: cumulative number of passengers who have arrived at the security queue by time $t$.
\item $d(t)$: cumulative number of passengers who have left the security by time $t$.
\item  $q(t)$: cumulative number of passengers whose flight have departed by time $t$.
\end{itemize}

We use $\hat{a}_{p}(t)$, $\hat{d}_{p}(t)$ to denote the respective cumulative arrival and departure functions under control policy $p$. We use total time savings (TTS), in passenger-hours, to quantify the benefits of control policy $p$:

\begin{align}
    TTS(p) = \int_{0}^{T} a(t) - d(t) dt - \int_{0}^{T} \hat{a}_{p}(t) - \hat{d}_{p}(t) dt \nonumber  \\ 
    \approx \sum_{t=1}^{T} (a(t) - d(t)) dt - \sum_{t=1}^{T} (\hat{a}_{p}(t) - \hat{d}_{p}(t)) dt
\end{align}

\begin{figure}[h!]
    \centering
    \includegraphics[width=0.5\textwidth]{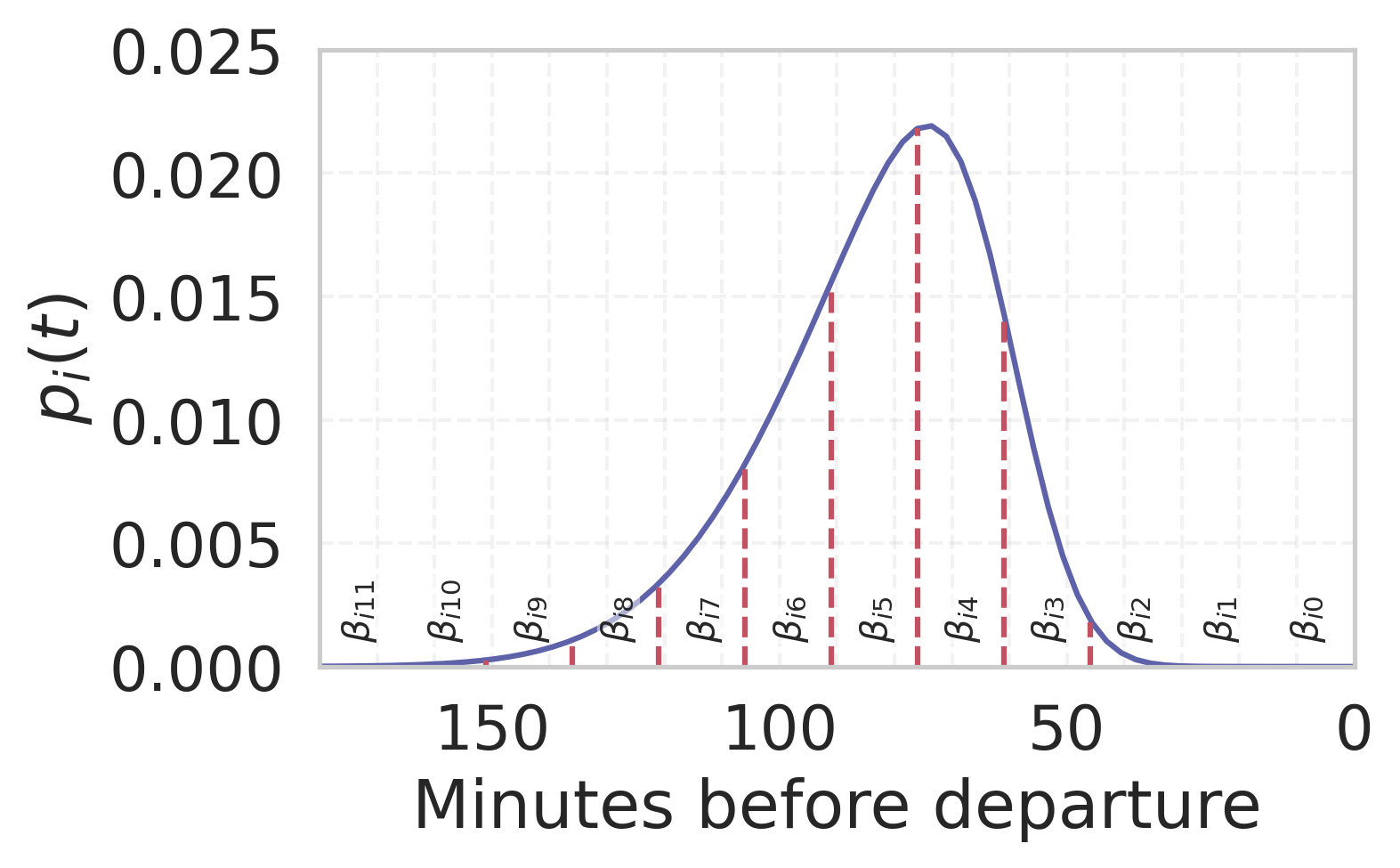} 
    \caption{Distribution of lead time for passengers on flight $i$. Lead time is defined as the difference in time between when a passenger arrives at security and when the flight departs. The above distribution is defined with a mean $\mu=64$, standard deviation $\sigma=30$, and a skewness parameter $\alpha=3$. The notation $\beta_{it}$ is the probability mass of passengers arriving in the $t^{\text{th}}$ slot prior to flight departure.}
    \label{fig:arrival_time_dist}
\end{figure}

We calculate the arrival time of the passenger at the security by subtracting the lead time of a passenger from the departure time of the person's flight. The lead time is defined as the time difference between when a passenger arrives at the security checkpoint and the departure time of the flight. We obtain the lead time distribution from the 2013 Airport Corporative Research Program (ACRP) report. The lead time distribution is characterized by a skew normal distribution with a mean of 64 minutes, standard deviation of 30 minutes, and a skewness parameter of 3 \cite{b1}. Figure \ref{fig:arrival_time_dist} shows the lead time distribution. The left-skewed distribution reflects the fact that more people choose to arrive at airports early rather than late. With the arrival time of passengers, we can compute the arrival curve $a(t)$. The corresponding departure curve $d(t)$ can then be calculated given a time-slot capacity.

\subsection{Deterministic Optimization Model}
\label{sec:det}

The deterministic formulation assumes that all passengers follow the recommended time of arrival in the security queue. We formulate the deterministic program as follows:

\begin{align}
    \text{min} \quad & \sum_{i \in F}\sum_{t \in T} x_{it} c_{it} d_{i} \nonumber \\
    \text{s.t.} \quad & \sum_{t\in T}x_{it} = 1 \quad \forall i \in F \label{cons:d1} \\
    \quad & \sum_{i\in F}x_{it}d_{i} \leq C_{t} \quad \forall t\in T \label{cons:d2}\\
    \quad & \sum_{s \in T \setminus Q_{i}} x_{is} = 0 \quad \forall i \in F \label{cons:d3} \\
    \quad & x_{it} \geq 0 \quad \forall i \in F, t\in T
    \label{cons:d4}
\end{align}

where $F$ is the set of departure flights and $T$ is the set of time slots to when passengers can be recommended to arrive. Decision variable $x_{it}$ represents the fraction of passengers on flight $i$ that is recommended to arrive at security checkpoints in time slot $t$. $c_{it}$ is the corresponding cost of such recommendation and $d_{i}$ is the number of passengers on flight $i$. Constraint (\ref{cons:d1}) ensures that all passengers receive a recommendation. Constraint (\ref{cons:d2}) ensures that the number of passengers arriving at security checkpoint in time slot $t$ does not exceed $C_{t}$, the capacity of the security system in time slot $t$. Constraint (\ref{cons:d3}) specifies the window of recommendation for passengers on flight $i$. We define the window of recommendation of passengers on flight $i$ as $Q_{i} = \{{t \in T | \text{dep}_{i}-M_{i} \leq t < \text{dep}_{i}}\}$, where $\text{dep}_{i}$ is the departure time of flight $i$, and $M_{i}$ is the maximum difference, in number of time slots, between the earliest recommended time slot and the departure time of the flight. We set $M_{i} = 16$, or 4 hours given 15-minute time slots. We set it to 4 hours because the support of the arrival time distribution of passengers on flight $i$, $p_{i}(t)$ is contained in $[\text{dep}_{i}-4\text{hrs}, \text{dep}_{i})$, as shown in Fig. \ref{fig:arrival_time_dist}. It is unlikely that passengers choose to arrive more than 4 hours before departure or after the flight has left. Therefore, it suffices to model arrivals within 4 hours prior to departure. This also helps reduce the number of free variables in the deterministic formulation and reduce the size of the inequality constraint matrices in the chance-constrained case that is discussed later. Constraint ($\ref{cons:d4}$) defines the domain of the decision variable $x_{it}$. 

We define the cost of a recommendation as the deviation of the recommended time from the mean lead time of 64 minutes (4 time slots). The cost of recommending passengers on flight $i$ to time slot $t$ is defined as:

\begin{align}
    c_{it} = \begin{cases}
        (\text{dep}_{i}-t)^{2} & \text{if } t \leq \text{dep}_{i}-\Bar{L} \\
        (\text{dep}_{i}-t) & \text{if }  \text{dep}_{i}-\Bar{L} < t \leq \text{dep}_{i}
    \end{cases}
\end{align}

where $\bar{L}$ is the mean lead time. Assigning passengers to time slots prior to the mean arrival time has a quadratic cost structure because we assume that passengers, in general, prefer to spend less time at the airport assuming they can make it to the flight. Assigning passengers to time slots after the mean arrival time, on the other hand, is assumed to incur a linear cost as it helps reduce the waiting time of passengers post-security. The notations used are summarized in Tab. \ref{tab:symbols} for reference.

\begin{table*}[h!]
    \centering
    \caption{Description of Symbols Used in the Formulation}
    \renewcommand{\arraystretch}{1.2}  
    \setlength{\tabcolsep}{3pt}        
    \small                             
    \begin{tabular}{|>{\raggedright\arraybackslash}m{5cm}|c|p{11cm}|}
        \hline
        \textbf{Category} & \textbf{Symbol} & \textbf{Description} \\
        \hline
        \multirow{3}{*}[0pt]{Sets} 
        & $F$ & Set of departure flights. \\
        & $T$ & Set of time slots for recommended security arrival. \\
        & $Q_{i}$ & Set of time slots available for flight $i$ recommendations. \\
        \hline
        Decision Variable & $x_{it}$ & Fraction of passengers on flight $i$ recommended to arrive in time slot $t$. \\
        \hline
         
        & $c_{it}$ & Cost of recommending passengers of flight $i$ to time slot $t$. \\
        & $d_{i}$ & Number of passengers on flight $i$. \\
       Parameters in both Models & $C_{t}$ & Security checkpoint capacity in time slot $t$. \\
        & $M_{i}$ & Maximum allowable difference between earliest recommendation and departure time. \\
        & $\text{dep}_{i}$ & Departure time of flight $i$. \\
        \hline
       
        & $\beta_{it}$ & Probability mass of passengers arriving in the $t^{\text{th}}$ slot prior to flight departure. \\
        & $\alpha_{it}$ & Compliance rate of passengers on flight $i$ for recommendation to slot $t$. \\
         Chance-Constrained Optimization Parameters & $\mu_{it}$ & Mean compliance rate of passengers on flight $i$ in slot $t$. \\
        & $\sigma_{it}$ & Standard deviation of the compliance rate $a_{it}$. \\
        & $\gamma$ & Reliability factor. \\
        & $p_{i}(t)$ & Probability density of the lead time of passengers on flight $i$. \\
        \hline
       
        & $a(t)$ & Cumulative number of passengers arriving at security queue by time $t$. \\
        & $d(t)$ & Cumulative number of passengers departing from security by time $t$. \\
        Queuing Model Notations  & $q(t)$ & Cumulative number of passengers whose flights have departed by time $t$. \\
        & $\hat{a}_{p}(t)$ & Cumulative arrival function under control policy $p$. \\
        & $\hat{d}_{p}(t)$ & Cumulative departure function under control policy $p$. \\
        & $TTS(p)$ & Total time savings in passenger-hours under policy $p$. \\
        \hline
    \end{tabular}
    \label{tab:symbols}
\end{table*}

\subsection{Chance-Constrained Optimization Model}

To account for varying level of passenger compliance rate, we reformulate the time slot recommendation problem as a second-order cone program with chance constraints. The chance constraints dictate with what probability the number of passenger arrivals at the security in each time slot is less than the security capacity. The intuition behind such a definition is that no waiting time is incurred when the number of passengers arriving in each time slot is less than the security capacity at the time.

We assume that a passenger can either comply with the recommendation and arrive in the recommended time slot or arrive according to the probability distribution of lead time as defined in Fig. \ref{fig:arrival_time_dist}. We define the compliance rate, $\alpha_{it}$, as the probability that passengers on the flight $i$ would accept the recommendation to arrive in slot $t$. We model compliance rate as a random variable indexed by both flight and time slot. Not only can there be day-to-day variation in compliance rate depending on the season or day of the week, but also flight-to-flight variance, as passengers on more business-oriented routes have less flexibility in their schedule. We assume a normal distribution in compliance rate:

\begin{equation}
\label{eq:alpha}
    \alpha_{it} \sim \mathbb{N}(\mu_{it}, \sigma_{it})
\end{equation}

The total number of passengers on flight $i$ arriving in time slot $t$ is the sum of passengers who follow the recommendation to arrive in time slot $t$ and those who do not comply with the assignment but arrive in time slot $t$ following the lead time distribution shown in Fig. \ref{fig:arrival_time_dist}. Therefore, we can express the number of passengers $y_{it}$ on flight $i$ who arrive at the security in time slot $t$ as follows:

\begin{equation}
    y_{it} = d_{i}x_{it}\alpha_{it} + \sum_{t\in Q_{i}} x_{it} (1-\alpha_{it})d_{i}\beta_{it}
\end{equation}

where $d_{i}$ is the number of seats or passengers on flight $i$ (which is the same as the number of passengers under our assumption of 100\% load factor) and $\beta_{it}$ is the probability that a passenger on flight $i$ arrives in the $t^{\text{th}}$ slot prior to the flight departure time when no control is in place, which is shown in Fig. \ref{fig:arrival_time_dist}. We can then express, in vector form, the total number of passengers arriving in time slot $t$ as:

\begin{align}
    y_{t} &= \vec{\alpha}^{\top}_{t}\mathbf{D}\vec{x}_{t} 
    - \sum_{t\in M}\vec{\alpha}_{t}\mathbf{B}_{t}\mathbf{D}\vec{x}_{t} + \sum_{t\in M}\mathbf{B}_{t}\mathbf{D}\vec{x}_{t}
\end{align}

where $\mathbf{D} \in \mathbb{R}^{N\times N}$ is a diagonal matrix in which the diagonal entries are the number of seats on flight $i$. $\mathbf{B_{t}} \in R^{N\times N}$ is also a diagonal matrix in which $\mathbf{B_{t}}ii = \beta_{it}$. $\vec{\alpha}_{t}$ is a vector that consists of the compliance rates for each flight. Hence, we can express the constraint that the number of passengers arriving is less than the security capacity in slot t, $C_{t}$, as:

\begin{align}
\label{eq:matrix}
    \xi_{t}^{\top}\mathbf{\Tilde{D}} \mathbf{A}_{t} & \bold{x}_t \leqslant C_{t} \quad \forall t \in T 
\end{align}


\begin{align}
\text { where } 
\xi_{t} & =\left[\begin{array}{ll}
\vec{\alpha_{t}}^{\top} & \overrightarrow{1}^{\top}
\end{array}\right]^{\top}, \quad 
    \xi_{t} \in \mathbb{R}^{2N\times 1} \\
\mathbf{\Tilde{D}} & =\left[\begin{array}{ccc}
\mathbf{D} & 0 \\
0 & \mathbf{D}
\end{array}\right], \quad 
\mathbf{\Tilde{D}} \in \mathbb{R}^{2N\times2N} \\
\mathbf{A}_{t} & =\left[\begin{array}{cccc}
I_N & -\mathbf{B}_{t} &\cdots & -\mathbf{B}_{t} \\
0_N & \mathbf{B}_{t} & \cdots & \mathbf{B}_{t}
\end{array}\right], \\  \mathbf{A}_{t} & \in \mathbb{R}^{2N \times (|M|+1)N} \nonumber\\
\bold{x}_t & =\left[\begin{array}{lllll}
\vec{x}_t^{\top} & \vec{x}_{1}^{\top} & \vec{x}_{2}^{\top} & \cdots & \vec{x}_{|M|}^{\top}
\end{array}\right]^{\top}, \\   \bold{x}_{t} & \in \mathbb{R}^{(|M|+1)N \times 1} \nonumber
\end{align}

We write the above matrices not just for arithmetic manipulation but because they carry intuitive meanings that allow our model to adapt to various implementations. Matrix $\Tilde{D}$ reflects the varying sizes of aircraft and the actual number of bookings on each flight. Matrix $A_{t}$ contains information on the arrival distributions of passengers at security on each flight. As the arrival distribution can vary across both time and flight, $A_{t}$ is capable of reflecting such variation. The vector $\bold{x}_{t}$ defines the potential time slots in which passengers on each flight can arrive. The notation $\vec{x}_{i}^{\top}$ in vector $\bold{x}_{t}$ denotes the corresponding decision variable $x_{it}$ that is 1,2,...,M slot prior to the departure time of flight $i$. $\xi_{t}$ is the uncertainty set that models the variability in compliance rate $\alpha_{it}$. Assuming the compliance rate follows a normal distribution, we define the uncertainty set as a multinomial normal distribution:

\begin{align}
    \xi_{t}\sim N\left(\mu_{t}, \Sigma_{t}\right)
\end{align}

where $\mu_{t}, \Sigma_{t}$ are respectively the mean and variance of $\xi_{t}$. Given a reliability factor of $\gamma$, which represents the probability that the constraint is violated, we then express the chance constraint as follows:

\begin{align}
\label{eq:refo}
\mathbb{P}_{\xi_{t} \sim N(\mu_{t}, \Sigma_{t})}\geq 1-\gamma
\end{align}

Let $\theta \sim$ $N(\mathbf{0}, \mathbf{I})$, we can reformulate the chance constraint (\ref{eq:refo}) to be

\begin{align}
\begin{aligned}[t]
\mathbb{P}\left(\mathbf{\Tilde{D}}\xi_{t}^{\top} \mathbf{A}_{t} \mathbf{x}_{t} \leq C_{t}\right) &= \\
\mathbb{P}\left(\theta \leq \frac{C_{t}-\mathbf{\Tilde{D}}\mu_{t}^{\top} \mathbf{A}_{t} \bold{x}_{t}}{\left\|\mathbf{x}_{t}^{\top} \mathbf{A}_{t}^{\top} \mathbf{\Tilde{D}}\boldsymbol{\Sigma}_{t}  \mathbf{\Tilde{D}} \mathbf{A}_{t} \bold{x}_{t}\right\|_2}\right) &\geq 1-\gamma
\end{aligned}
\end{align}

which can take a second-order cone form of

\begin{align}
    {\left\|\mathbf{x}_{t}^{\top} \mathbf{A}_{t}^{\top} \mathbf{\Tilde{D}}\boldsymbol{\Sigma}_{t}  \mathbf{\Tilde{D}} \mathbf{A}_{t} \bold{x}_{t}\right\|_2}\leq \frac{1}{\Phi^{-1}(1-\gamma)}\left(C_{t}-\mathbf{D}\mu^{\top}_{t} \mathbf{A}_{t} \mathbf{x}_{t}\right)
\end{align}

The chance-constrained optimization problem can be defined as follows:

\begin{align}
    \text{min} \quad & \sum_{i \in F}\sum_{t \in T} x_{it} c_{it} d_{i} \nonumber \\
    \text{s.t.} \quad & \sum_{t\in T}x_{it} = 1 \quad \forall i \in F \label{cons:r1} \\
    \quad &     {\left\|\mathbf{x}_{t}^{\top} \mathbf{A}_{t}^{\top} \mathbf{\Tilde{D}}\boldsymbol{\Sigma}_{t}  \mathbf{\Tilde{D}} \mathbf{A}_{t} \bold{x}_{t}\right\|_2}\leq \nonumber \\ \quad &\frac{1}{\Phi^{-1}(1-\gamma)}\left(C_{t}-\mathbf{D}\mu^{\top}_{t} \mathbf{A}_{t} \mathbf{x}_{t}\right) \quad \forall t\in T \label{cons:r2}\\
    \quad & \sum_{s \in T \setminus Q_{i}} x_{is} = 0 \quad \forall i \in F \label{cons:r3} \\
    \quad & x_{it} \geq 0 \quad \forall i \in F, t\in T
    \label{cons:r4}
\end{align}

The chance-constrained optimization takes the same format as the deterministic case discussed in Sec. \ref{sec:det} except that we replace the deterministic constraints that the security capacity cannot be exceeded with chance constraints (\ref{cons:r2}). We retain the original definition of the cost function.

\begin{figure}[h!]
    \centering
    \includegraphics[width=0.5\textwidth]{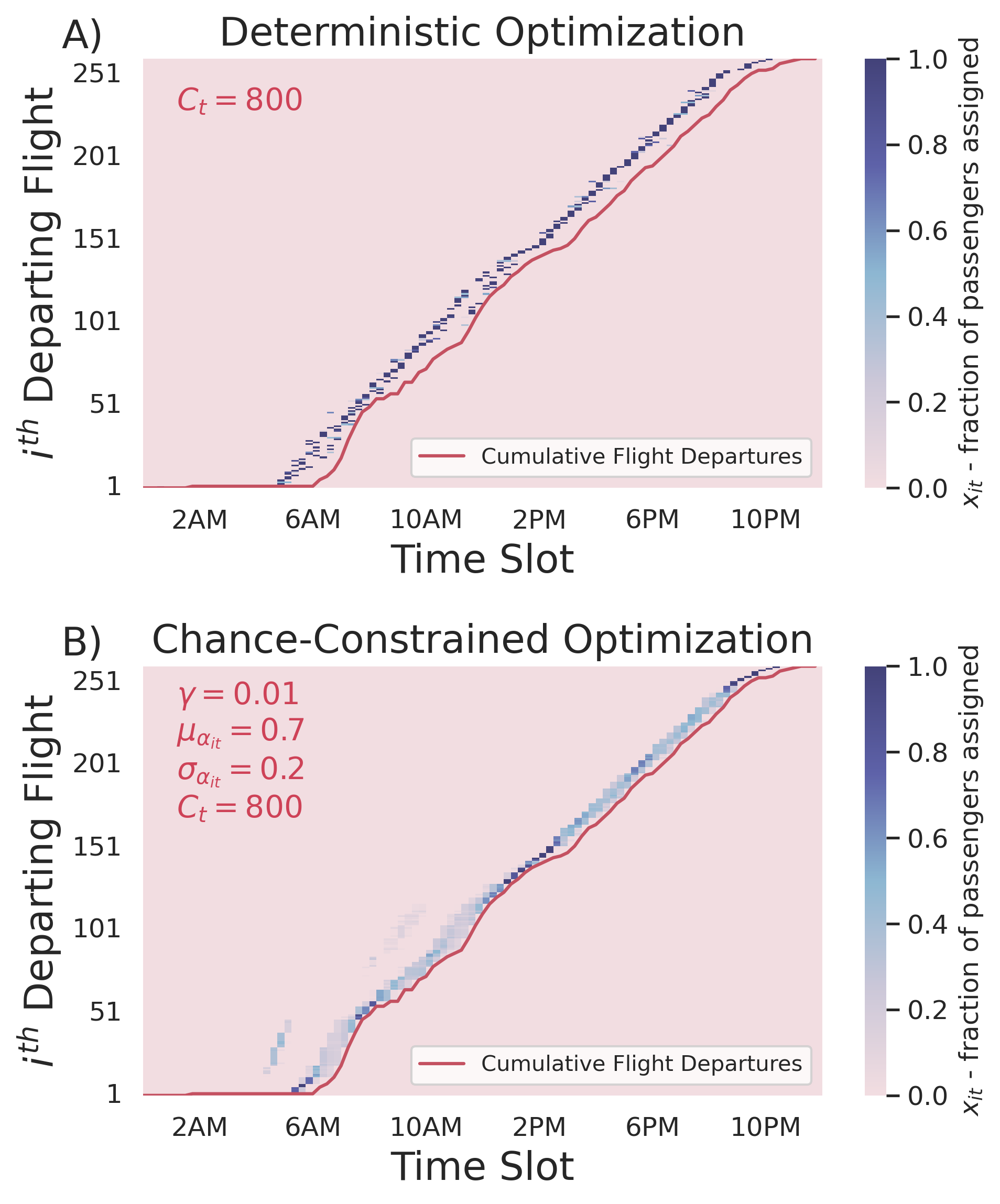} 
    \caption{Solution to the two optimization problems. A) Solution to the deterministic optimization problem. B) Solution to the chance-constrained optimization problem assuming a average compliance rate, $\alpha_{it}$ of 0.7 with a standard deviation of 0.2. The reliability factor, $\gamma$, is set to 0.01.}
    \label{fig:assignment}
\end{figure}
\section{Numeric Experiment}
\label{sec:exp}

\begin{figure*}[h!]
  \centering
  \includegraphics[width=1\linewidth]{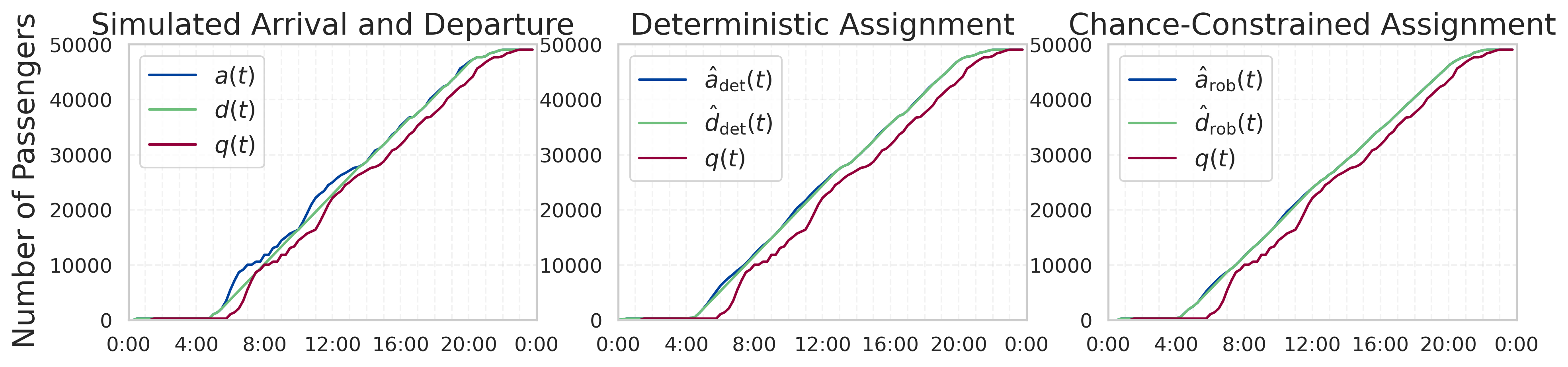}
     \caption{Queuing diagram of security checkpoint. We assume a capacity of 800 passengers per time slot}
  \label{fig:queue}
\end{figure*}

We used the departure flight schedule from Terminal 1 at Josep Tarradellas Barcelona-El Prat Airport (BCN) of May 9th, 2023 to demonstrate the effectiveness of our model in reducing waiting time at security. The schedule comprises 260 departure flights with a total of 49,034 seats. Consistent with the formulation, we set time slots to 15-minute intervals and $M$ to 16 (4 hours).

Figure \ref{fig:assignment} illustrates the optimal solution to both the deterministic and the chance-constrained optimization problem. In both cases, the security capacity is assumed to be 800 passengers per time slot (15 minutes). While the deterministic program gives a discrete solution for the majority of flights, the solution to the chance-constrained optimization problem is much smoother, meaning that it spreads passengers on the same flight over a number of time slots. Especially in peak periods such as early morning and late afternoon, we see passengers being referred to a number of neighboring slots. During the less-busy period, the chance-constrained optimization gives a solution similar to that of the deterministic problem, as it is less likely that the capacity of security would be exceeded given the fewer departure flights during those periods. 

It is also worth pointing out that a fraction of passengers on some flights is recommended to time slots much earlier than the departure time in the chance-constrained optimization. This recommendation occurs when a rapid increase in the influx of passengers is observed. Between 6am and 8am, as well as 12pm and 2pm, we see the highest flight departure rate at BCN. Therefore, to ensure that queues do not build up at security, the optimization recommends a fraction of passengers on the flight in those time periods to time slots much earlier than their flight departure time. An alternative solution would be to recommend that all passengers arrive slightly earlier. However, such a solution would result in a higher objective function value as the number of passengers whose recommendation deviates from the mean arrival time is much greater. 

To quantify the benefits of the recommendations, we calculated the total time savings (TTS) of the two control policies. We use algorithm \ref{algo:1} to simulate the arrival of passengers at security given a control policy $p$. The algorithm samples the compliance rate from Eq. (\ref{eq:alpha}) and compute the realized distribution of arrival time based on the control policy. We then generate the arrival time of each passenger using the realized distribution of arrival time. For the base case with no control in place, we simply generate the arrival time of the passengers based on the probability distribution shown in Fig. \ref{fig:arrival_time_dist}. 

\begin{figure*}[h!]
  \centering
  \includegraphics[width=1\linewidth]{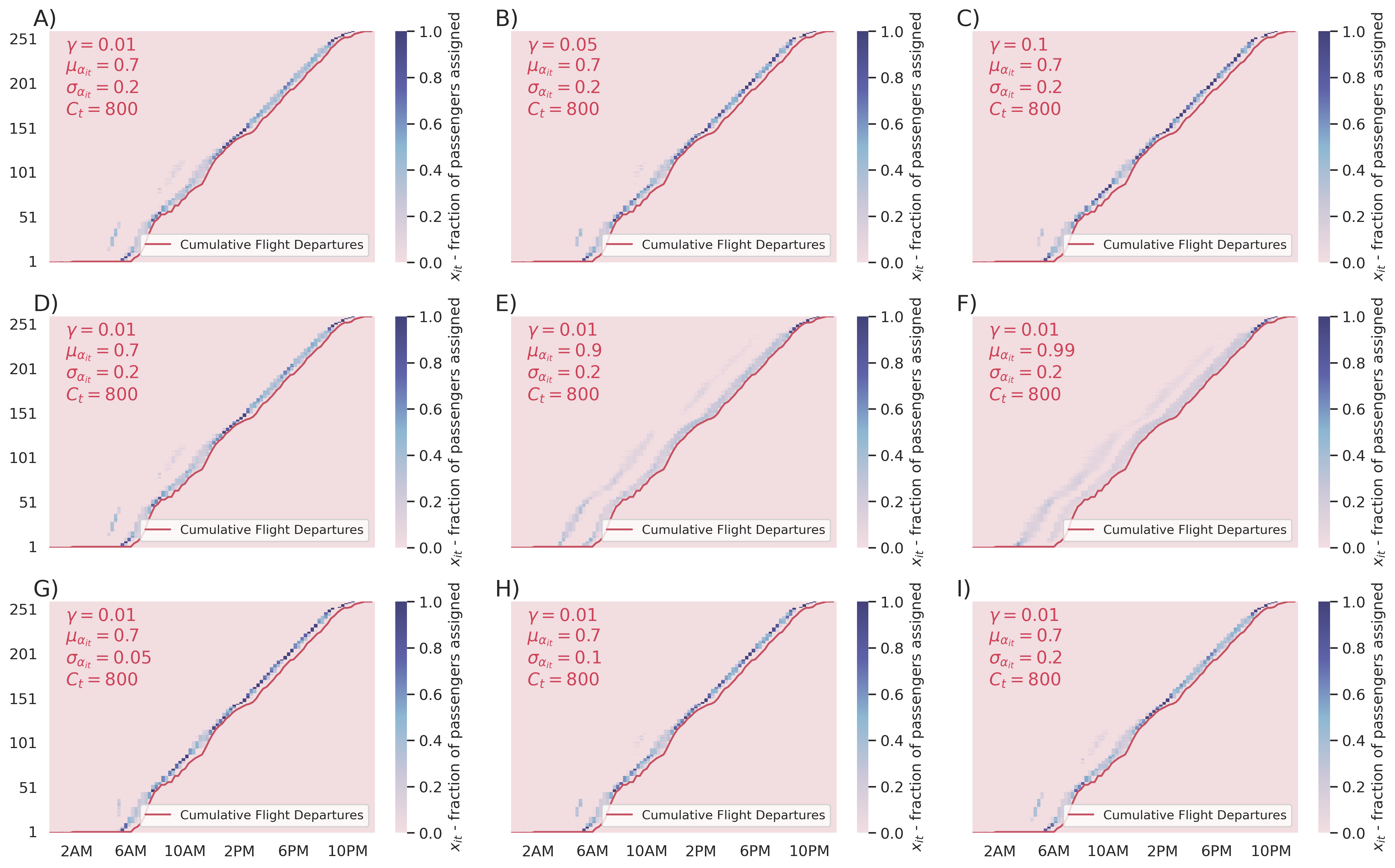}
     \caption{Sensitivity analysis. We vary one parameter while keeping the rest fixed for comparison. (A)-(C) shows the solutions when varying $\gamma$. (D)-(F) shows the solutions when varying $\mu$. (G)-(I) shows the solutions when varying $\sigma$}
  \label{fig:sensitivity}
\end{figure*}

\begin{algorithm}[h!]
	\caption{Generating Passenger Arrivals at Security}
	\begin{algorithmic}

        \State  $a$ = [] \Comment{Initialize an empty list of passenger arrival time}
        
        \For{$i \in F$}
            \State Initialize $\hat{p}_{i}$ = [] \Comment{Initialize an empty array for the probability mass of arrival time of passengers on flight $i$ under a control policy}
            \For{$t \in T$}
                \State Sample $\hat{\alpha}_{it}$ from $a_{it} \sim \mathbb{N}(\mu_{it}, \sigma_{it})$
                \State $\hat{p}_{i}$[t] = $x_{it} \hat{\alpha}_{it}$ \Comment{Compute the arrival density for those who comply with the recommendation}

                \For{$s\in Q_{i}$}
                    \State $\hat{p}[s] += x_{it} \hat{\alpha}_{it} \beta_{is}$
                    \Comment{Compute the arrival density for those who do not comply with the recommendation}
                \EndFor

            \EndFor
            \State Sample $d_{i}$ numbers from the probability mass function defined by $\hat{p}_{i}(t)$ and add the arrival time to list $a$
        \EndFor

	\end{algorithmic}
    \label{algo:1}
\end{algorithm}

Figure \ref{fig:queue} shows the cumulative arrivals and departures in the three scenarios: no control, deterministic control, and robust control. We observe two periods, early morning and noon, when queues build up at security, which correspond to the time period with the most flight departures. We see a significant reduction in queue length and waiting time under both the deterministic control and the robust control scheme. The TTSs of the robust and deterministic control are 8,161 and 6,009 passenger-hours, representing a 85\% and 65\% decrease in total waiting time, respectively. Seemingly, we get the counterintuitive result that accounting for variance in fact decreases the total waiting time at security; however, it is important to point out that the total waiting time at security is not directly captured in the objective function. The objective function measures the deviation of the recommended arrival time from the passengers' preferred time, which is guaranteed to be higher under the chance-constrained optimization model compared to the deterministic case. The total amount of time passengers spend at the airport can be higher or lower compared to the deterministic case, depending on the flight schedule and the solution. Therefore, accounting for stochasticity in passengers' arrival indeed leads to an increase in the objective function value but also to a decrease in waiting time at security, as the arrival of passengers are more spread out. The difference between the two TTS is attributed to passenger non-compliance and the variance in compliance rate. This difference highlights the importance of considering passenger compliance in reducing total waiting time. 

Furthermore, we see that in the case without control, there exists a time in which $q_(t)$ exceeds the $d(t)$ curve, which means that some passengers would have missed their flight in that case. However, the control policy derived from both deterministic and chance-constrained optimization does not lead to passengers missing their flights.

\section{Sensitivity Analysis}
\label{sec:sen_ana}

To showcase the flexibility of the chance-constrained model in modeling different passengers and scenarios, we present a sensitivity analysis of the parameters that characterize passenger compliance and the degree to which chance constraints are enforced. The sub-figures in each row of Fig. \ref{fig:sensitivity} isolate the effect of varying one parameter while keeping the others fixed, enabling a structured comparison of their influence on compliance behavior and system reliability. 

Sub-plots (A-C) illustrate how changes in the reliability factor $\gamma$ affect the distribution of recommendations. As $\gamma$ increases, the system enforces stricter probability bounds on constraint violations, improving robustness but reducing feasible solutions. Therefore, we see that with the highest $\gamma=0.1$, few passengers are assigned slots much earlier than the departure time compared to the assignment in (A), where $\gamma=0.01$. However, this also means that with higher $\gamma$, we are likely to observe a longer waiting time at security.

Sub-plots (D-F) analyze variations in the mean compliance rate $\alpha$, where an increase in mean compliance leads to a systematic change in the distribution, revealing a trade-off between system flexibility and reliability. Interestingly enough, we see that with a higher compliance rate, a larger fraction of passengers are recommended to arrive in earlier time slots. In fact, we see a divergence in passengers on the same flight. This phenomenon can be explained by the fact that when the compliance rate is low, many more passengers would arrive at the security following the arrival distribution in Fig. \ref{fig:arrival_time_dist}. A fraction of the non-complying passengers will arrive at security checkpoints early at zero cost, which is the reason why fewer passengers are recommended to arrive early in (D). 

Finally, sub-plots (G-I) explore the implication of the standard deviation of $\alpha$, showing that a greater variability in compliance introduces greater fluctuations in reliability. As expected, a higher variance in passenger compliance leads to a solution that is more conservative in the recommendation. We see that many more passengers are recommended to arrive early in (I) when compared to (G). Overall, Figure~\ref{fig:sensitivity} highlights the interplay between these three parameters, emphasizing the balance required between robustness, compliance, and uncertainty in the chance-constrained framework.

\section{Conclusion}

In this paper, we proposed a chance-constrained optimization model that recommends the arrival time at the security for passengers. The model takes into account the variation in passenger compliance, an important human factor. By converting uncertainty in passenger compliance into chance constraints, we ensure with a confidence factor of $1-\gamma$ that queues will not build up at security check points. The results of our numeric experiment demonstrate the effectiveness of the control policy derived from the chance-constrained optimization, as a reduction in waiting time of 85\% is achieved. We further show that such a reduction is 30\% more compared to the deterministic case, which does not consider passenger compliance or the variability in it. It highlights the importance of incorporating passenger compliance in the design of security checkpoint control policies.

For future research, one direction would be to incorporate varying compliance rates across different flights and time periods. Passengers are more likely to comply with recommendations that are closer to the mean arrival time given the lead time distribution. In addition, the structure of the cost function can be improved by a more detailed estimation of the preferences of the passengers. Although we assume that passengers spending less time at the airport leads to lower cost, it can be argued that it also leads passengers to develop anxiety about missing the flights, resulting in disutility. Therefore, it is important to examine the structure of the cost function so that it not only best reflects the preferences of the passengers, but also aligns the commercial interests of the airport with those of the passengers.

It would also be exciting if we had the opportunity to conduct field experiments in airports and test the effectiveness of the model in real-life scenarios. With implementation comes data, which will enable us to derive estimates of compliance rates. Estimation can be performed online and a feedback control loop can be established to incorporate changes in the estimation, enhancing both robustness and responsiveness. Another potential direction is to incorporate the supply side of the equation into the picture. We can use multi-objective optimization models to obtain control policies that not only recommend passengers arrival time but also determine the service rates of the security system through decisions such as staffing.

\vspace{12pt}


\begin{thebibliography}{00}
\bibitem{bounce} C. Cody, "Airport Wait Times 2024", Bounce, Oct., 2024, \url{https://bounce.com/blog/airport-wait-times-2024}.

\bibitem{b7} D. Bertsimas and M. Sim, "The Price of Robustness," \textit{Operations Research}, vol. 52, no. 1, pp. 35-53, 2004. https://doi.org/10.1287/opre.1030.0065.

\bibitem{brun} B. Alexis, F. Eric, A. Sameer, D. Daniel, "Schedule Optimization and Staff Allocation for Airport Security Checkpoints using Guided Simulated Annealing and Integer Linear Programming", \textit{Journal of Air Transport Management}, vol. 124, 2025.

\bibitem{b9} E. Perez, L. Taunton, and J. A. Sefair, "A Simulation-Optimization Approach to Improve the Allocation of Security Screening Resources in Airport Terminal Checkpoints," \textit{2021 Winter Simulation Conference (WSC)}, pp. 1–11, Dec. 2021. https://doi.org/10.1109/WSC52266.2021.9715421.

\bibitem{gilliam} G. Ronald, "An application of queuing theory to airport passenger security", \textit{Interfaces}, vol. 9, no. 4, pp. 117-123, 1979. https://doi.org/10.1287/inte.9.4.117.

\bibitem{iata} International Air Transportation Association, "Global Air Passenger Reaches Record High in 2024", Jan., 2025, \url{https://www.iata.org/en/pressroom/2025-releases/2025-01-30-01/}.

\bibitem{chen} J. Chen, L. Chen, and D. Sun, "Air traffic flow management under uncertainty using chance-constrained optimization", \textit{Transportation Research Part B: Methodological}, vol. 102, pp. 124-141, 2017.

\bibitem{gmp} K. Mun Hwan, P. Jin Woo, and C. Yu Jin, "A Study on the Effects of Waiting Time for Airport Security Screening Service on Passengers’ Emotional Responses and Airport Image", \textit{Sustainability}, vol. 12, no. 24, pp. 10634-10640, 2020. https://doi.org/10.3390/su122410634.

\bibitem{b1} K. Patrick, H. Robert, F. Eric, and H. Belinda, "Guidelines for Preparing Peak Period and Operational Profiles (2013)," National Academies of Sciences, Engineering, and Medicine, 2013. https://doi.org/10.17226/22647.

\bibitem{lange} L. Robert, S. Ilya, and R. Bo, "Virtual queuing at airport security lanes", \textit{European Jorunal of Operations Research}, vol. 225, no. 1, pp. 153-165, 2013.

\bibitem{fox} M. Daniel, "Here are the airports with the longest security wait times during the holidays: Study", fox news, Nov., 2024, \url{https://www.livenowfox.com/news/airports-longest-security-wait-times-holidays}.

\bibitem{marshall} M. Zachary, M. John, G. Adam, P. Caleb, and D. Luigi, "Expediting airport security queues through advance lane assignment", \textit{J Transp Secur}, vol. 15, issue 3-4, pp. 245-262, 2022.

\bibitem{eu}{Qsensor, \textit{The 11 Worst European Airports for Security Queues - June Edition}, Qsensor, Jun., 2023.}

\bibitem{b6} R. Burkard, M. Dell’Amico, and S. Martello, \textit{Assignment Problems}. Society for Industrial and Applied Mathematics, 2009.

\bibitem{b3} S. Alodhaibi, R. L. Burdett, and P. K. D. V. Yarlagadda, "Impact of passenger-arrival patterns in outbound processes of airports," \textit{Procedia Manufacturing}, vol. 30, pp. 323-330, 2019. https://doi.org/10.1016/j.promfg.2019.02.046.

\bibitem{c1} S. Gary, "One of Every Seven Travelers Miss Their Flights Because
of Long Airport Security Lines", Forbes News, Jun., 2018.

\bibitem{scozzaro} S. Geoffrey, M.M. Miguel, D. Daniel, and M. Catherine, "Simulation-Optimisation-Based Decision Support System for Managing Airport Security Resources", EUROSIM 2023, pp. 140-145, 2024, \url{https://doi.org/10.1007/978-3-031-68438-8_11}.

\bibitem{b5} Transportation Research Board, "Airport Passenger-Related Processing Rates Guidebook," Airport Cooperative Research Program (ACRP) Report 23, 2009. \url{https://crp.trb.org/acrp0715/wp-content/themes/acrp-child/documents/029/original/ACRP_23_Airport_Passenger-Related_Processing_Rates_Guidebook.pdf}.
 
\bibitem{b4} Transportation Research Board, "Passenger Value of Time, Benefit-Cost Analysis and Airport Capital Investment Decisions," Airport Cooperative Research Program (ACRP) Report 107, 2015. \url{https://crp.trb.org/acrp0715/wp-content/themes/acrp-child/documents/186/original/acrp_wo22.pdf}.

\bibitem{wang} W. Ruiting, K. Patrick, Z. Teng, S. Jairo, V. Aashrith, B. Hoseinali, and M. Scott, "Robust routing for a mixed fleet of heavy-duty trucks with pickup and delivery under energy consumption uncertainty", \textit{Applied Energy}, vol. 368, 2024.

\bibitem{b2} Z. A. Marshall, J. H. Mott, A. J. Gottwald, C. A. Patrick, and L. R. Dy, "Expediting airport security queues through advanced lane assignment," \textit{Journal of Transportation Security}, vol. 15, pp. 245-262, 2022. https://doi.org/10.1007/s12198-022-00247-9.


\end{thebibliography}
\end{document}